\documentstyle[floats,twocolumn,epsf,prl,aps]{revtex} 

\title{ Sign Reversals of ac Magnetoconductance in Isolated Quantum Dots} 

\author{Yves Noat $^1$, H\'el\`ene Bouchiat$^1$, Bertrand Reulet$^1$ and
Dominique Mailly$^{2}$}

\address{$^1$Laboratoire de Physique des Solides, Associ\'e au CNRS, B\^at 510,
Universit\'e Paris--Sud, 91405, Orsay, France.\\ $^2$C.N.R.S Laboratoire de
Microstructures et de Micro\'electronique, 196, Avenue Ravera, 92220, Bagneux,
France\\ \parbox{14cm}{\medskip\rm\small%
 We have measured the
electromagnetic response of micron-size isolated mesoscopic GaAs/GaAlAs square
dots down to temperature $T=16mK$, by coupling them to an electromagnetic
micro-resonator. Both dissipative and non dissipative responses  exhibit a
large magnetic field  dependent quantum correction, with a characteristic flux
scale which corresponds to a flux quantum in a dot.  The real (dissipative)
magnetoconductance changes sign as a function of frequency for low enough
density of electrons. The signal observed at frequency below the mean level
spacing $\Delta$, corresponds to a negative magnetoconductance, which is
opposite to the weak localization seen in connected systems, and becomes
positive at higher frequency. We propose an interpretation of this phenomenon
in relation to fundamental properties of energy level spacing statistics in the
dots. }}

\begin{document}

\maketitle

There are several ways of measuring the electrical conductance of a metallic
sample. The most common one is to drive a current through it and measure the
voltage drop between the connected wires. But, it is also possible to measure
the conductance without making any connection to the sample. Capacitance and
inductance measurements at finite frequency provide such a way. When a metallic
sample of typical size $a$ is inserted into a coil, it acquires a magnetic
moment $m\simeq Ia^2$, where $I$ is the eddy current induced in the sample.
This changes the impedance of the coil by a quantity: $\delta Z(\omega) \simeq
{\cal L}\omega \chi_m(\omega)/V$  where $V$ is the volume of the coil of
inductance ${\cal L}$ and $\chi_m$ is the magnetic susceptibility of the sample
 related to the conductance $G_m$ , the ratio between  $I$ and
the induced electromotive force by \cite{LL}:

\begin{equation}
	\chi_m(\omega)= i\omega \mu_0 G_m(\omega) a^4  \label{gm}
\end{equation}

Similarly the admittance change of a capacitance $\cal C$ due to the insertion
of a metallic sample reads: $\delta Z^{-1}(\omega)\simeq{\cal
C}\omega\chi_e(\omega)/V$ where $V$ is the volume of the capacitor and
$\chi_e(\omega)$, the electrical polarisability of the sample, is related to its
effective conductance $G_e$ (defined as ratio between polarization current and
unscreened voltage drop through the sample), by \cite{def}:

\begin{equation} \chi_e(\omega)=\frac { G_e(\omega)a^2}{i\omega \epsilon _0 }
\end{equation}
	 In the classical limit at low frequency (such that the size of the sample is
much smaller than the skin depth) $G_m$ is real and identical to the Drude
conductance $G_D=\sigma a$ within a numerical constant which depends on the geometry of
the sample. On the other hand due to screening of the electric field inside the
metal \cite{SI}, and for $\omega < \sigma/\epsilon_0$, $G_e$ is dominated by its imaginary part: Im$G_e= \omega a \epsilon_0$.
Its real part is equal to: 

\begin{equation}{\rm Re} G_e= (\epsilon_0 \omega a)^2 /G_D \label{eqge}
\end{equation}

In a mesoscopic sample, which length is smaller than the
electronic phase coherence length, it was shown theoretically \cite{BIL,TR88,RB
th,KRGB,ZS} that the conductance strongly  depends on the way it is measured.
More precisely, it is very sensitive to the coupling between the mesoscopic
sample and the macroscopic classical apparatus. When a sample is connected
with wires, (that is to say when the coupling is strong), relative quantum
corrections on the conductance are of the order of $1/g$  where $g
=G_D/(e^2/h)$\cite{AAS}.  This result concerning the dissipative
components of the conductance are  unchanged for unconnected samples provided
that the energy level spectrum is continuous \cite{KRGB} i.e. the typical level
width $\gamma$ is  much larger than the mean level spacing $\Delta$. Note
however that the quantum corrections of Re$ G_e$ and Re$G_m$ are
opposite in sign. Furthermore Im$ G_m$ is expected to be finite and directly
related to persistent currents and more generally orbital magnetism in the
sample \cite{BIL,TR88,RB th,efetov}. At the same time  Im$G_e$ acquires a
magnetic field dependence related to quantum corrections to the electrical
polarisability \cite{pola}.

On the other hand, if the discrete spectrum limit $\gamma\ll\Delta$ is
achieved, quantum coherence of the sample  gives rise to giant
magnetoconductance \cite{RB th,KRGB,ZS} whose sign and amplitude are determined
by the order relation between the relevant energy scales:
$\hbar\omega,\gamma,\Delta,k_B T$. Preliminary experiments 
 \cite{RRB exp} on an array of Aharonov-Bohm  rings coupled to a
resonator, have shown that the investigation of ac conductance on isolated
samples in the discrete spectrum limit is indeed possible. 
\begin{figure} 
\[\epsfbox{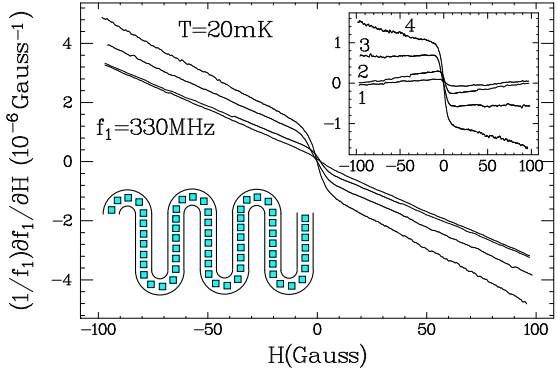}\]
 \caption{Field derivative of
resonance frequency for the first harmonics of the resonator, the different
curves correspond to successive illuminations of the sample.Inset, bottom: schematic picture of the resonator and the square dots. Inset, top : data obtained
after substraction of the empty resonator contribution.  \label{figdfh}}
\end{figure}

In this letter, we report the measurement of the electromagnetic response of an
array of mesoscopic square dots made from GaAs/GaAlAs 2D electron gas at
frequencies both below and above energy level spacing. Sign reversals  of
magnetoconductance were observed which can be interpreted  in relation to the
sensitivity to magnetic field of  level spacings  distribution  in the dot, 
according to random matrix theory.

As in the previous experiment\cite{RRB exp}, the mesoscopic samples are coupled
to an electromagnetic micro-resonator, made of two Niobium wires ($d=4\mu$m
spaced, $1\mu$m width and $20$cm long) deposited in a meandering geometry on a
sapphire substrate. The  superconducting material is used to reduce
 losses and therefore get a high quality factor ($80000$ without
any sample). The  resonance conditions are:
  $L=\frac{n \lambda}{2}$, where $n=1,2,3...$ and $L$ is the length of the
superconducting wires and $\lambda$ the electromagnetic wavelength. The
fundamental resonance frequency  is of the order of $330$MHz. This experiment
allows us to detect extremely small variations of frequency and quality factor: 
$ \frac{\delta f}{f} \sim \delta (Q^{-1}) \sim 10^{-9}$.
In the presence of mesoscopic samples deposited on the top, (with a spacer
consisting of a $1\mu$m thick mylar sheet),  the characteristics of the
resonator are modified. The dissipation  in the samples affects the quality
factor, whereas the non dissipative response
affects the resonance frequency:

\begin{equation} \frac{\delta f}{f} = -k N_s  \chi'(\omega) , \delta(Q^{-1}) =
k N_s \chi''(\omega) \end{equation} where
$\chi(\omega)=\chi_e(\omega)+\chi_m(\omega)$ is the average complex
susceptibility of a dot \cite{def}, $N_s$ is the number of samples and $k$ is a
geometrical coupling coefficient  which is of the order of $1/Ld^2$, where $d$
is the distance between the two Niobium wires. The sample studied in this
experiment, was an array of $10^5$ disconnected square dots  etched, using
electron beam lithography techniques, in  a semiconductor heterojunction
GaAs/GaAlAs. The electron concentration  of the two dimensional electron gas
(2DEG) before etching was $n_e =3.10^{11}{\rm cm}^{-2}$ and elastic mean free
path $l_e=10\mu$m. This electronic density was strongly depressed in etched
samples but could be recovered with illumination. For this purpose an
electro-luminescent diode was placed next to the resonator, allowing to change
gradually the number of electrons in the squares by transient illumination of
the samples (between 30s and 2mn). The measurements were always done at least
one hour after the illumination was stopped, in order to ensure good stability
of the samples. The size of the squares $a = 1.5\mu$m was estimated from weak
localization experiments on connected samples made together with the squares.
The transport in the samples can thus be considered as ballistic. The mean
level spacing in a square $\Delta = 2\frac{\hbar^2 \pi}{m^*a^2}$, ($m^*$ is the
effective electron mass) is independent of the electronic density. It is of the
order of $38$mK which corresponds to a frequency of $760$MHz. Resonator and
diode are enclosed in a metallic box, protecting from electromagnetic
noise, and  cooled with a dilution fridge down to $16$mK.

\begin{figure}
 \[\epsfbox{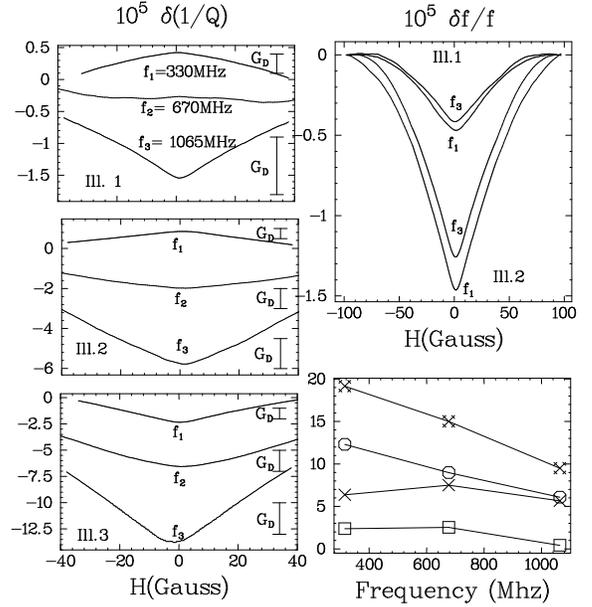}\] 
\caption{Field dependence of real
and imaginary susceptibilities for several resonance frequencies and
illuminations. The amplitude of the signal corresponding to the Drude value
$G_D$ is estimated for each illumination assuming that  $\delta \chi'(f_1)$ is
proportional to $G_D$ and that illumination 4 corresponds to saturation i.e. a
density of 6 $10^{11}$ el./cm$^2 $. Bottom right: frequency dependence of $\delta \partial f/ \partial H$ in units of $10^{-7}$ Gauss$^{-1}$.  \label{figimreGH}} \end{figure}	

We have measured the magnetic field dependence of the resonance frequency and
quality factor for several harmonics ($f_1=330$MHz, $f_2=670$MHz,
$f_3=1065$MHz) and after successive illuminations. The dc magnetic field was
modulated at low frequency ($ \sim 30$Hz) with a small coil close to the
sample. This allows measurement of derivatives of resonance frequency (see
fig.\ref{figdfh}) and quality factor with respect to magnetic field. The signal
of the samples consists in step-like features on the field dependence of
$\partial f/\partial h \propto -\partial\chi'(H)/\partial H$ and $\partial Q^{-1}/\partial
h\propto \partial \chi''(H)/ \partial H$ whose width $\simeq 10$ Gauss corresponds
approximatively to half a flux quantum in a square. In the case of $\partial
f/\partial h$  the contribution of the empty resonator, linear in magnetic
field, has to be subtracted from the signal. After integration,  the
magnetoconductance $\delta G(H) \propto \delta \chi(H)= \chi(H) -\chi(0)$ of the samples can be deduced from these measurements (see
fig.\ref{figimreGH}). We observe that both real and imaginary components present the striking triangular shape observed in experiments
on connected ballistic dots of similar geometries which is believed to be
related to the classical integrability of these
systems\cite{chang}.

$\delta\chi'(H)$ and $\delta \chi''(H)$ are found to have very different
behaviors as a function of frequency. $\delta\chi'(H)$ is always positive, its
amplitude  drastically increases  with the first successive illuminations which
correspond to an increasing number of electrons, and then tends to saturate.
For a fixed number of electrons it is nearly independent of
frequency\cite{price}. On the other hand $\delta \chi''(H)$ is observed to be
negative for the first harmonics and small electronic density and becomes positive
with increasing frequency and electronic density. This sign change of the magnetoconductance is the most striking result of our work.

\begin{figure} \[\epsfbox{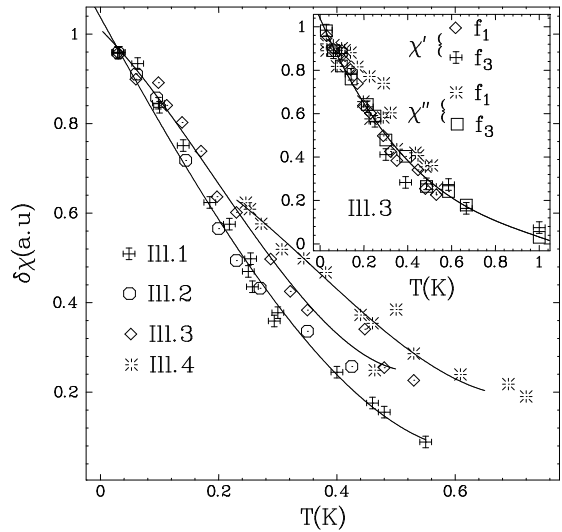}\] \caption{Rescaled, temperature dependences of
$\delta \chi'$ for the first harmonics and various number of electrons
corresponding to successive illuminations (continuous lines are only guides for the eyes). Inset: temperature dependence of
$\delta \chi'$ and $\delta \chi'' $for the first and third harmonics.  
\label{figgt}} \end{figure}

 The temperature dependence of the signals were also measured. The temperature
decay depicted in fig.\ref{figgt} is found to be independent of frequency and
nearly identical for $\delta\chi'$ and  $\delta\chi''$. It cannot be fitted with an exponential as was the case for the ring experiment.  The characteristic
temperature scale  of this  decay slightly increases with the number of
electrons in the dots.

In the following we want to compare our results with theoretical predictions on
orbital magnetism and polarisability.  The orbital susceptibility of ballistic
squares has been calculated by several authors, it is found to be paramagnetic
in zero field, in agreement with experimental results on the dc magnetization
of GaAs/GaAlAs squares\cite{Levy}. Our results on the other hand indicate
unambiguously low field diamagnetism, just like the previous ac experiment on an ensemble
of rings\cite{RRB exp}. We note however that the amplitude of $\delta \chi'(H)$
measured at saturation  correspond to an average orbital susceptibility of
the order of the theoretical prediction: $ga^2 \mu_0e^2/m^*$ \cite{ulmo}. In the case
of an electric coupling, the signal can be related to the average
magneto-polarisability of the squares. This quantity has been recently
calculated\cite{pola},  its sign (which corresponds to an increase of
polarisability with magnetic field) could explain our experimental results on
$\delta \chi'(H)$.  Nevertheless, according to these predictions, the
magneto-polarisability is proportional to $1/g$ and also to the Thomas-Fermi
screening length. Thus it should decrease with the number of electrons, which
is in complete contradiction with what we measured. So, the understanding of our
measurements on $\delta\chi'(H)$ deserves further theoretical investigations.

We now consider the dissipative part of the response. The low frequency
classical absorption of a metallic grain in an ac electric field has been
shown\cite{SI,ZS} to be described by the effective conductance $G_e$ (see
expression (\ref{eqge})). This result initially derived for 3D systems can easily
be extended to 2D samples provided that the electric field lies in their plane.
The ratio between ``magnetic'' losses described by $G_m$ and ``electric''
losses in the resonator  is then simply given by: $r=(Z_0G_D)^2$, where
$Z_0=\sqrt{ \frac{\mu_0}{\epsilon_0}}= 377\Omega $ is the vacuum 
impedance. In our case $r$ varies between 8 and $400$ depending
on the electronic density in the dots . Therefore, we conclude that the
dominant contribution to $\chi''$ comes from $G_m$. Using
expression (\ref{gm}) relating $G_m$ to the measured quantity $\delta \chi''(H)$,
it is possible to express the magnetoconductance measured for several
frequencies and number of electrons in units of $G_D$ as shown in
fig.\ref{figimreGH}. Note that in most cases the magneto-conductance is of the
order of $G_D$ as expected for a discrete level system.

The field dependence of the dissipation of Aharonov-Bohm rings in a time
dependent flux has been  extensively studied \cite{BIL,TR88,RB th,KRGB}. It is
straightforward to extend these results to singly connected geometries and
estimate expected $\delta G_m(H)$ in our case. Let us recall the existence of
two different contributions to $\delta G_m(H)$  in the limit where
$\gamma\ll\Delta $ and $T>\Delta$: $\delta G_{off}$  and
$\delta G_{di}$. 
$\delta G_{off}$ related to inter-level transitions is expressed in terms of the the level spacing distribution function $R(s)$ 
 which obeys universal rules of random matrix theory\cite{metha,GE}.
It corresponds for $H=0$  to Gaussian Orthogonal Ensemble (GOE) and for
sufficiently large field, in order that time reversal symmetry is broken, to Gaussian Unitary Ensemble (GUE).
 
\begin{equation} 
\delta G_{off}(H)=G_D \int \frac{\gamma (R_{GUE}(s)-R_{GOE}(s))}{(\hbar\omega -s)^2 +\gamma^2} ds
\end{equation}
On the other hand $\delta G_{di}$ is related to the relaxation of persistent currents which  always yields to positive magnetoconductance:

\begin{equation} 
 \delta G_{di}(H)=  G_{di}(GUE) - G_{di}(GOE)= G_D \frac{\gamma \Delta} {\omega
^2+\gamma ^2} \label{eqg2} 
\end{equation}  

\begin{figure}
 \[\epsfbox{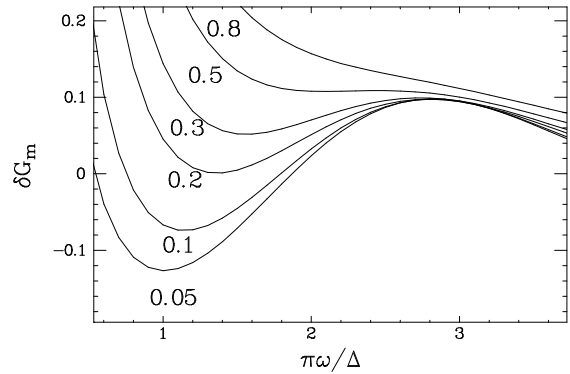}\] 
\caption{ Calculated frequency dependence of $\delta
G_m$ for different values of the ratio $\gamma /\Delta$. \label{dgtheo}}
\end{figure}

In the limit where  $\gamma\ll\omega $ the magneto-conductance is dominated by:
$\delta G_{off}(H)\sim G_D (R_{GUE}(\omega)-R_{GOE}(\omega))$. The
magnetoconductance is negative at  $\omega<\Delta$ and changes sign at
frequency of the order of $\Delta$ as depicted in fig.\ref{dgtheo}. We believe
that the squares after illumination 1 correspond to this situation.
This change of sign is however expected to disappear when $\gamma $ increases
such as the condition: $\delta G_2 > |\delta G_1|$ is realized. We suggest that
illuminations  3 and 4 correspond to this last  situation. The increase of
$\gamma$ for successive illuminations is necessary to explain our results.
It cannot be due to the contribution of electron-electron interactions (whose
contribution decreases with increasing electron density\cite{AG}) but possibly
to the  increase of losses in the etched GaAs substrate. We indeed observed a substantial decrease of Q after each illumination.  The
electromagnetic noise related to these losses is  expected to contribute
substantially to the level broadening\cite{kamenev,BR98}. Theory \cite{RB
th,KRGB} also predicts, for canonical ensemble, changes of sign of the
magnetoconductance with  temperature at $T<\Delta$ which we have not
observed, possibly because of lack of experimental points in the regime where $T\ll \Delta$.

In conclusion these results show that ac conductance measurements on isolated
samples reveal indeed new physical features. Giant magnetoconductance has been
observed. Sign changes on the real part are in agreement with the pioneer
predictions of Gorkov and Eliashberg \cite{GE} on the sensitivity of the energy spectrum
to time reversal breaking by  magnetic field.  The imaginary part whose
amplitude is of the same order of magnitude than the real one is in principle
related to the orbital magnetism in the dots. However its sign is not yet understood. 

We aknowledge fruitfull discussions with B. Altshuler, Y. Gefen, B.Schlovski and P.Walker.  


\begin{thebibliography}{99}
\vspace{-1cm}
\bibitem{LL}  L.Landau and E. Lifchitz, "Electrodynamics of continuous media",
Mir publishers, (1969)  

\bibitem{def} Within this notations the magnetization and the electric
polarization of a dot read: $ m(\omega) = \chi_m H(\omega) $ and
$p(\omega)=\epsilon_0 \chi_e E(\omega)$, where $H$ and $E$ are respectively the
ac magnetic and electric field.

 \bibitem{SI} U. Sivan and Y. Imry,  Phys. Rev. B, {\bf 35}, 6074, (1987)
 

\bibitem{BIL} M. B\"uttiker, Y. Imry and R. Landauer, Physics Letters
{\bf 96A}, 365, (1983); R. Landauer and M. B\"uttiker, Phys Rev. Lett. {\bf
54}, 2049 (1985).


\bibitem{TR88} N. Trivedi and D.A. Browne, Phys.Rev. B{\bf 38},9581,(1988). 
\bibitem{RB th} B. Reulet and H. Bouchiat, Phys.Rev. B{\bf 50},2259,(1994).

\bibitem{KRGB} A. Kamenev, B. Reulet, H. Bouchiat and Y. Gefen, Europhys.
Lett.  {\bf 28}, 391, (1994).


\bibitem{ZS} F. Zhou, B. Spivak, N. Taniguchi and B. L. Altshuler,  Phys.
Rev. Lett {\bf 77}, 1958, (1996). 

\bibitem{AAS} B. L. Altshuler, A. G. Aronov and B. Z. Spivak, Pis'ma Zh. Eksp.
Teor. Fiz. {\bf 33}, 101 (1981) [JETP Lett. {\bf 33}, 94 (1981)], D. Y. Sharvin and Y. V. Sharvin, Pis'ma Zh. Eksp. Teor. Fiz.
{\bf 34}, 285 (1981) [JETP Lett. {\bf 34}, 272 (1981)]

\bibitem{efetov} K. B. Efetof, Phys. Rev. Lett {\bf 66}, 2794, (1991). 

\bibitem{pola} Y. Noat, B. Reulet and  H. Bouchiat, Europhys. Lett. {\bf
36}(9), 701 (1996) . Y.Blanter and Mirlin (preprint)

\bibitem{RRB exp} B. Reulet, M. Ramin, H. Bouchiat and D. Mailly,  Phys. Rev.
Lett {\bf 75}, 124 (1995). 



\bibitem{chang} A. M. Chang, H. U. Baranger, L. N. Pfeiffer and K. W. West,
 Phys. Rev. Lett {\bf 73}, 2111 (1994). 

\bibitem{price}  Note that in our experiments the amplitude of the imaginary
magnetoconductance is always of the same order of magnitude than the real one.
This finding is fundamentally different from results obtained on ac
measurements on a connected system where imaginary conductance could only be
detected for frequencies larger than the Thouless Energy. J. B. Pieper and J.
C. Price,  Phys. Rev. Lett. {\bf 72}, 3586 (1994)  J. B. Pieper, J.
Price and J. Martinis,  Phys. Rev. B, {\bf 45}, 3857 (1992).

\bibitem{Levy} L. L\'evy, H. Reich, L. Pfeiffer and K. West,  Physica B,
{\bf 189}, 204 (1993).

\bibitem{ulmo} Y. Gefen, D.Braun and G.Montambaux, Phys. Rev. Lett. {\bf 73}, 
154  (1994);D.Ullmo, K.Richter and R.Jalabert, Phys. Rev. Lett. {\bf 74},  383
 (1995).

\bibitem{metha}  M. L. Mehta, {\it  Random Matrices and the Statistical Theory
of Energy Levels}. New York : Academic Press 1967.

\bibitem{GE} L.P. Gor'kov, G.M. Eliashberg, Sov. Phys.-JETP {\bf 21}, 940
(1965);  B. I. Shklovskii, Pis'ma Zh. Eksp. Teor. Fiz. {\bf 36}, 287 (1982)  [
JETP Lett. {\bf 36}, 352 (1982)].

\bibitem{AG} U. Sivan, Y. Imry and A. G. Aronov, Europhys. Lett. {\bf
28}, 115 (1994); B. L. Altshuler, Y. Gefen, A. Kamenev and S. L. Levitov,
 Phys. Rev. Lett, {\bf 78}, 2803 (1997). 

\bibitem{kamenev} A.Kamenev and Y.Gefen "Correlated Fermions and transport in
mesoscopic structures", T. Martin, G. Montambaux and J. Trƒn Thanh Vƒn eds,
(1996)

\bibitem{BR98} B.Reulet et al. unpublished. 

\end{thebibliography}
\end{document}